# Direct Plan Comparison of RapidArc and CyberKnife for Spine Stereotactic Body Radiation Therapy

Young Eun Choi, PhD, Jungwon Kwak, PhD, Si Yeol Song, MD, Eun Kyung Choi, MD, Seung Do Ahn, MD, Byungchul Cho, PhD*

*Department of Radiation Oncology, Asan Medical Center, University of Ulsan College of Medicine, Seoul, Korea*

**\*Corresponding author:** Byungchul Cho, PhD, Department of Radiation Oncology

Asan Medical Center, University of Ulsan College of Medicine, 88 Olympic-ro 43-gil, Songpa-gu, Seoul 138-736, Korea

Tel.: (822) 3010 4437; Fax: (822) 3203 4437; E-mail: bcho@amc.seoul.kr

**Abstract**

We compared the treatment planning performance of RapidArc (RA) vs. CyberKnife (CK) for spinal stereotactic body radiation therapy (SBRT). Ten patients with spinal lesions who had been treated with CK were re-planned with RA, which consisted of two complete arcs. Computed tomography (CT) and volumetric dose data of CK, generated using the Multiplan (Accuray) treatment planning system (TPS) and the Ray-Trace algorithm, were imported to Varian Eclipse TPS in Dicom format, and the data were compared with the RA plan using analytical anisotropic algorithm (AAA) dose calculation. The optimized dose priorities for both CK and RA plans were similar for all patients. The highest priority was to provide enough dose coverage to the planned target volume (PTV) while limiting the maximum dose to the spinal cord. Plan quality was evaluated with respect to PTV coverage, conformity index (CI), high-dose spillage, intermediate-dose spillage ($R_{50\%}$ and $D_{2cm}$), and maximum dose to the spinal cord, which are criteria recommended by RTOG 0631 spine and 0915 lung SBRT protocols.

The mean CI ± SD values of PTV were 1.11 ± 0.03 and 1.17 ± 0.10 for RA and CK ($p$ = 0.02), respectively. On average, the maximum dose delivered to the spinal cord in CK plans was approximately 11.6% higher than in RA plans, and this difference was statistically significant ($p < 0.001$). High-dose spillages were 0.86% and 2.26% for RA and CK ($p$ = 0.203), respectively. Intermediate-dose spillage characterized by $D_{2cm}$ was lower for RA than for CK; however, $R_{50\%}$ was not statistically different. Even though both systems can create highly conformal volumetric dose distributions, the current study shows that RA demonstrates lower high- and intermediate-dose spillage than CK. Therefore, RA plans for spinal SBRT may be superior to CK.

# I. Introduction

Up to 40% of cancer patients develop spinal metastasis [1] , and radiation therapy (RT) is a well-established treatment for spinal tumors. Palliative RT with conventional techniques can effectively control pain [2, 3], but spinal metastases often recur because spinal cord tolerance limits the administered dose [4]. Accordingly, spinal stereotactic body radiotherapy (SBRT), which delivers a high radiation dose in a single or limited number of fractions within a small target volume, is the best approach for increasing local control because it minimizes the dose to the adjacent spinal cord while maximizing the dose delivered to the tumor region. This technique therefore can also significantly reduce radiation-induced myelopathy [5, 6].

Spinal SBRT can be performed using either a linear accelerator-based multileaf collimator (MLC) or a frameless real-time tumor tracking-supported CyberKnife (CK). CK can deliver multiple non-isocentric beams by 6 MV linear accelerator to the desired target. This system uses two ceiling-mounted diagnostic kV imagers to monitor the patient's position and track the tumor throughout treatment [7-9]. Volumetric-modulated arc radiotherapy (VMAT) can deliver highly conformal radiation doses in 1–5 fractions under image guidance by gantry rotation of the linear accelerator [10-12]. However, since these two modalities have different characteristics, they can result in different dose distributions. To address the different dosimetric characteristics of linear accelerator-based SBRT and CK, we directly compared the RapidArc (RA) plan with the CK plan for spinal SBRT, and evaluated plan quality using the conformity index (CI) and several SBRT-specific dose gradient indices.

## II. Materials and Methods

*Treatment planning*

Ten patients with spinal lesions (one patient with C-spine, four patients with T-spine, and five patients with L-spine), treated using CK at Asan Medical Center from January to June 2014, were retrospectively analyzed. The total prescribed dose was 18–26 Gy for 3–4 fractions, and the average dose per fraction was 6.5 Gy (range = 6–8 Gy). The CK plans were generated with a planning Computed tomography (CT) images (1.25 mm thick slices) using the Multiplan (Accuray Inc., Sunnyvale, CA) treatment planning system (TPS). One to three different sizes of fixed collimators and between 150 and 247 beams were used to fully cover the target. Dose distribution was calculated by the Ray-Trace algorithm.

The CK plans were re-planned with RA consisting of two complete arcs. For RA plan, the CT data with the calculated dose volume were imported into the Varian Eclipse TPS (version 10.0) in Dicom format and then compared with those of RA plans. The isocenter was set to the center of the planning target volume (PTV). Two complete arcs (clockwise rotation from 181–179° with the collimator at 30°, and counterclockwise rotation from 179–181° with the collimator at 330°) were applied to all RA plans. A single arc is consisted of 178 control points, roughly every 1°, to cover the PTV by moving the gantry at the maximum speed to minimize the treatment time. The RA plans were calculated using Eclipse (version 10.0.28) analytical anisotropic algorithm (AAA) dose calculation with a grid size of 1.25 mm. 10-MV flattening-filter-free beams of a TrueBeam accelerator equipped with a high-definition 120 multileaf collimator (Varian Medical System, Palo Alto, CA) were used. Optimization constraints were set to produce dose distributions that met the criteria of the RTOG protocols [13, 14]. The optimized dose priorities were similar for all cases. Providing enough dose coverage to the PTVs and limiting the maximum dose to the spinal cord were the highest priorities. Normalization was required for 97–100% of the isodose line to ensure the prescribed dose covered 95% of the PTV. Great care was needed to optimize the treatment plan: the dose constraint was < 0% for the spinal cord volume receiving 16 Gy ($V_{0\%}$ <16 Gy). The monitor units per fraction and the estimated delivery time were summarized for each patient in Table 1. The monitor units per fraction ranged from 1943 to 5971 for RA plan and from 8785 to 21872 for CK plan, respectively. The average estimated treatment times were 4.1 ±1.4 and 59.8 ±4.1 minutes for RA and CK, respectively.

Table 1. Summary of monitor units and beam-on time per fraction of CyberKnife and RapidArc for ten patients

| Patient No. | Dose per fraction [Gy] | No. of beams for CK | Monitor Units /Fraction | | Treatment Time/ Fraction [min] | |
|---|---|---|---|---|---|---|
| | | | RA | CK | RA | CK |
| 1 | 8 | 181 | 5971 | 15915 | 6.5 | 66 |
| 2 | 6 | 172 | 3830 | 10550 | 4.2 | 58 |
| 3 | 6.5 | 163 | 3897 | 11698 | 4.3 | 58 |
| 4 | 6.5 | 154 | 3492 | 11181 | 3.8 | 58 |
| 5 | 6.5 | 170 | 5933 | 11773 | 6.5 | 60 |
| 6 | 6.5 | 177 | 3678 | 12521 | 4.0 | 63 |
| 7 | 6 | 195 | 3111 | 11450 | 3.4 | 62 |
| 8 | 6.5 | 150 | 2846 | 8785 | 3.1 | 52 |
| 9 | 6.5 | 186 | 3304 | 12843 | 3.6 | 64 |
| 10 | 8 | 247 | 1943 | 21872 | 2.1 | 57 |
| Mean ± SD | | 179.5 ± 27.5 | 3801 ± 1268 | 12859 ± 3651 | 4.1 ±1.4 | 59.8 ± 4.1 |

*Dosimetric parameters*

The parameters used to evaluate plan quality included target volume coverage, CI, high-dose spillage, intermediate-dose spillage ($R_{50\%}$ and $D_{2cm}$), and maximum dose administered to the spinal cord, as recommended by the RTOG 0631 spine and 0915 lung SBRT protocols [13-15]. CI was defined as the ratio of PTV covered by the prescribed dose (PTV $V_{prescribed\ dose}$) to PTV. Acceptable conformity was defined as <1.2, according to the RTOG 0915 protocol [14]. High-dose spillage was calculated as the ratio of volume outside the PTV that received >105% of the prescribed dose to PTV volume (V [$V_{105\%}$ - PTV] / [PTV]). As shown in Figure 1, intermediate-dose spillage was defined as the fall-off gradient located outside of the PTV. Intermediate-dose spillages were expressed as $R_{50\%}$ (volume that received 50% of the prescribed dose/PTV volume) and $D_{2cm}$ (maximum dose in terms of the percentage of the prescribed dose at 2 cm beyond PTV in any direction). The accepted values for $R_{50\%}$ and $D_{2cm}$ varied depending on the PTV volume [14].

To evaluate the characteristics of each modality with respect to dose distribution, the dose distributions of the two plans were directly subtracted (RA - CK), from which dose difference patterns were observed when the RA plan was higher and lower doses than the CK plan, separately.

*Statistical analysis*

First, the Kolmogorov–Smirnov test [16] was used to determine whether the dosimetric parameters used for RA and CK plans were normally distributed or not. CI and high-dose spillage were analyzed using the Wilcoxon signed-rank test [16], and intermediate-dose spillage and the spinal cord dose were analyzed using the paired *t*-test (SPSS Statistics, version 18, IBM, NY). In this study, $p < 0.05$ was considered statistically significant.

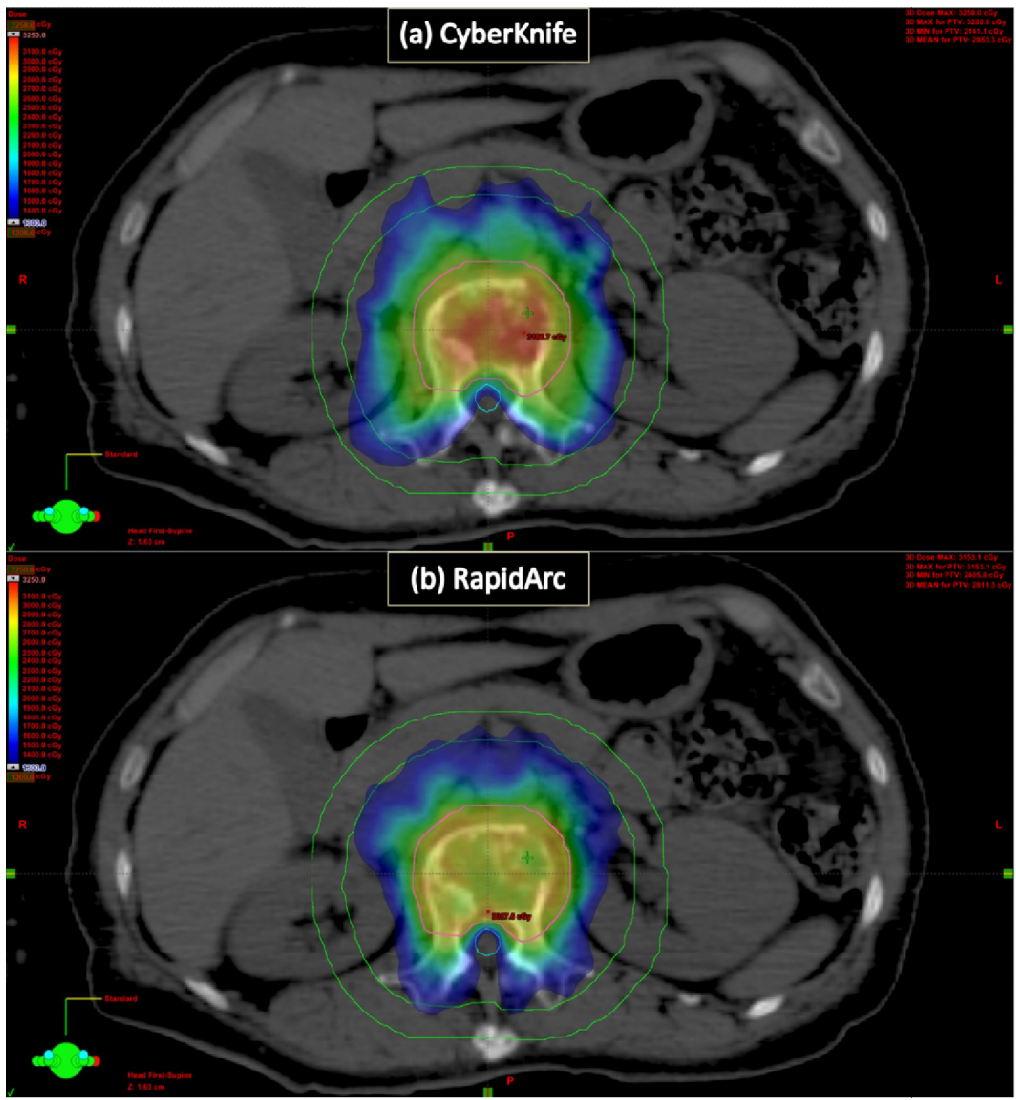

**Figure 1.** Intermediate-dose spillage was defined as the dose fall-off gradient beyond PTV ($R_{50\%}$ and $D_{2cm}$). Fifty percent of the prescribed dose was distributed, and the volume was measured to calculate $R_{50\%}$. For $D_{2cm}$, a 1 cm thick ring at 2 cm from the surface of the PTV in any direction was defined as the shell structure, and its maximum dose was determined from the dose-volume histogram (DVH). $R_{50\%}$ was (a) 5.20 for CK and (b) 3.64 for RA. $D_{2cm}$ was (a) 79.8% for CK and (b) 55.4% for RA. The PTV of the spinal SBRT-treated patients at L1 was 55.8 cc. According to the RTOG 0813 and 0915 protocols, $R_{50\%}$ and $D_{2cm}$ should not be >4.0 and 62.0%, respectively. RA satisfied these criteria, but CK resulted in minor violations.

## III. Results

*Target volume coverage*

The dose distribution within the PTV was assessed by evaluating the maximum, minimum, and mean doses. The CK plans consistently exhibited high mean and maximum doses in the PTV, and the volumes of the 100% prescribed doses are summarized in Table 2. The maximum dose was within the PTV for all plans. Figure 2 shows a representative DVH of the RA and CK plans. The RA plan produced better PTV coverage. Figure 3 compares the dose distribution and PTV coverage of the RA vs. CK plans in three directions. For the RA plan, the entire dose was homogeneously distributed within the PTV. The mean CI ± SD of PTV for the RA and CK plans were 1.11 ± 0.03 (range = 1.07–1.16) and 1.17 ± 0.10 (range = 1.09–1.42), respectively (Table 3). Three CK plans demonstrated CI values >1.2, whereas all RA plans had CI values <1.2; however, the two plans did not demonstrate statistically significant differences in conformity ($p = 0.074$).

*Dose gradient*

High-dose spillages were 0.86 ± 1.30% for RA and 2.26 ± 4.21% for CK ($p = 0.203$), as shown in Table 2. Intermediate-dose spillages, as characterized by $D_{2cm}$, were lower for RA than for CK ($p = 0.001$); however, $R_{50\%}$ did not show a statistically significant difference.

*Dose difference*

When subtracting the dose distribution of CK from RA, (RA - CK), the trace of the oblique incidence by the non-coplanar beams was within the range of the negative values (shown in pink); however, the pattern of the coplanar beams using both complete arcs was greater than 0 (shown in green) (Figure 4). The CK plan resulted in a higher dose than the RA plan inside and near the PTV, whereas the dose for the RA plan was distributed throughout the entire body.

*Normal tissue sparing*

DVHs were used to compare the maximum doses to the spinal cord (Figure 2). Eight CK plans exhibited higher doses than that of RA plans, and the average maximum dose to the spinal cord using CK was approximately 11.6% higher than when using RA (Figure 2 and Table 3). This difference was statistically significant ($p = 0.014$). Figure 5 compares the dose profiles for Patient #3. Vertical and horizontal lines were drawn through the center of the spinal cord for each plan. The vertical and horizontal dose profiles for RA (red line), CK (blue line), and differences between the two plans (pink line) are shown in Figures 5 (c–d). The spinal cord is highlighted by the blue box. The maximum doses to the spinal cord were 1957.6 and 1532.0 cGy for CK and RA, respectively. The RA plan resulted in steep dose gradients just outside the target, and thus resulted in a smaller dose to the spinal cord.

**Table 2. Summary of target dosimetric parameters for ten patients using CyberKnife and RapidArc.**

| Patient No. | PTV [cc] | $D_{min}$ [cGy] | | $D_{mean}$ [cGy] | | $D_{max}$ [cGy] | | $V_{100\%}$ [cc] | |
|---|---|---|---|---|---|---|---|---|---|
| | | RA | CK | RA | CK | RA | CK | RA | CK |
| 1 | 44.4 | 2125.1 | 1856.4 | 2584.2 | 2762.0 | 2866.0 | 3076.9 | 49.1 | 62.9 |
| 2 | 93.1 | 1699.0 | 1637.0 | 1934.6 | 1989.9 | 2149.8 | 2222.2 | 99.4 | 104.9 |
| 3 | 110.0 | 2449.4 | 2010.3 | 2789.6 | 2901.7 | 3141.9 | 3291.1 | 120.5 | 124.2 |
| 4 | 55.8 | 2405.8 | 2141.1 | 2811.3 | 2853.3 | 3153.1 | 3250.0 | 61.1 | 61.0 |
| 5 | 106.6 | 2217.4 | 1792.8 | 2816.4 | 2879.8 | 3195.8 | 3209.9 | 123.8 | 129.6 |
| 6 | 99.4 | 2439.6 | 1887.4 | 2819.5 | 2888.4 | 3141.4 | 3209.9 | 113.5 | 120.3 |
| 7 | 55.3 | 2142.6 | 1479.1 | 2560.8 | 2584.0 | 2870.9 | 2963.0 | 64.1 | 60.1 |
| 8 | 124.1 | 2503.8 | 2122.7 | 2788.6 | 2846.0 | 3031.4 | 3291.1 | 135.7 | 140.4 |
| 9 | 50.6 | 2162.3 | 1366.0 | 2798.6 | 2807.4 | 3071.2 | 3250.0 | 57.2 | 60.3 |
| 10 | 12.4 | 1772.1 | 1360.5 | 2610.0 | 2651.9 | 2901.2 | 3116.9 | 13.4 | 13.8 |
| Average | | 2191.7 | 1765.3 | 2651.4 | 2716.4 | 2952.3 | 3088.1 | 83.8 | 87.8 |
| SD | | 278.5 | 293.4 | 272.6 | 276.1 | 307.8 | 321.6 | 40.2 | 41.5 |

PTV, planning target volume; $V_{100\%}$, the volumes of PTV receiving the 100% prescribed dose.

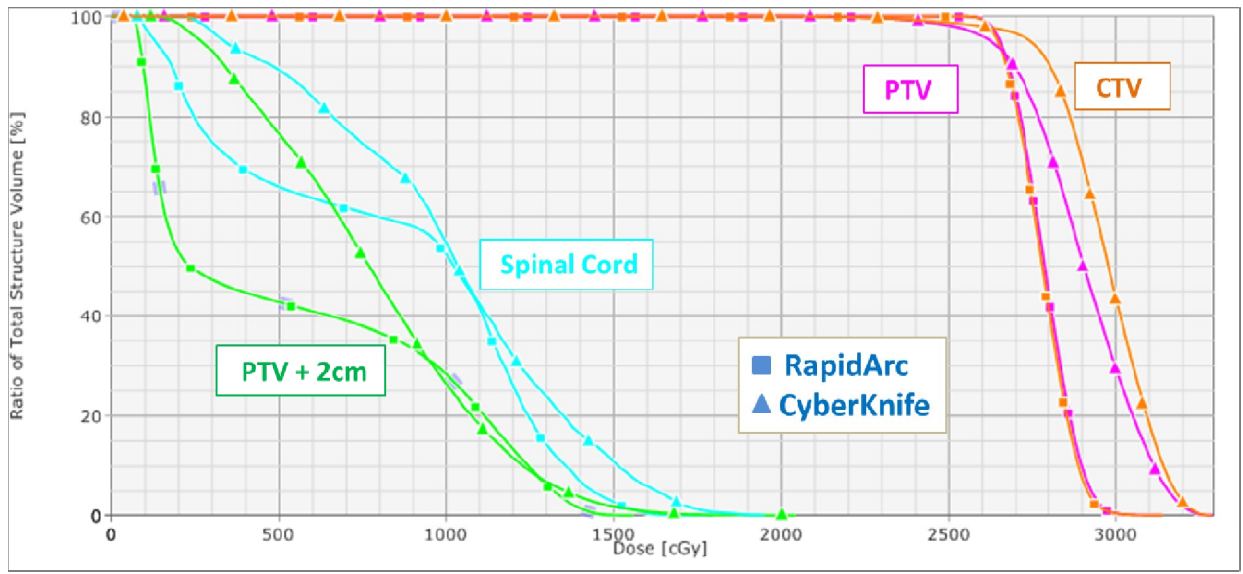

**Figure 2. DVH for RapidArc (■) and CyberKnife (▲) in Patient #3 (L2 spinal lesion).** PTV (pink), CTV (orange), spinal cord (cyan), and PTV + 2 cm (green) are shown. The RA plan achieved higher PTV coverage than the CK plan. The maximum doses to the spinal cord and the shell located at 2 cm around the PTV were 17.9 Gy and 15.6 Gy for RA, and 19.5 Gy and 20.4 Gy for CK, respectively.

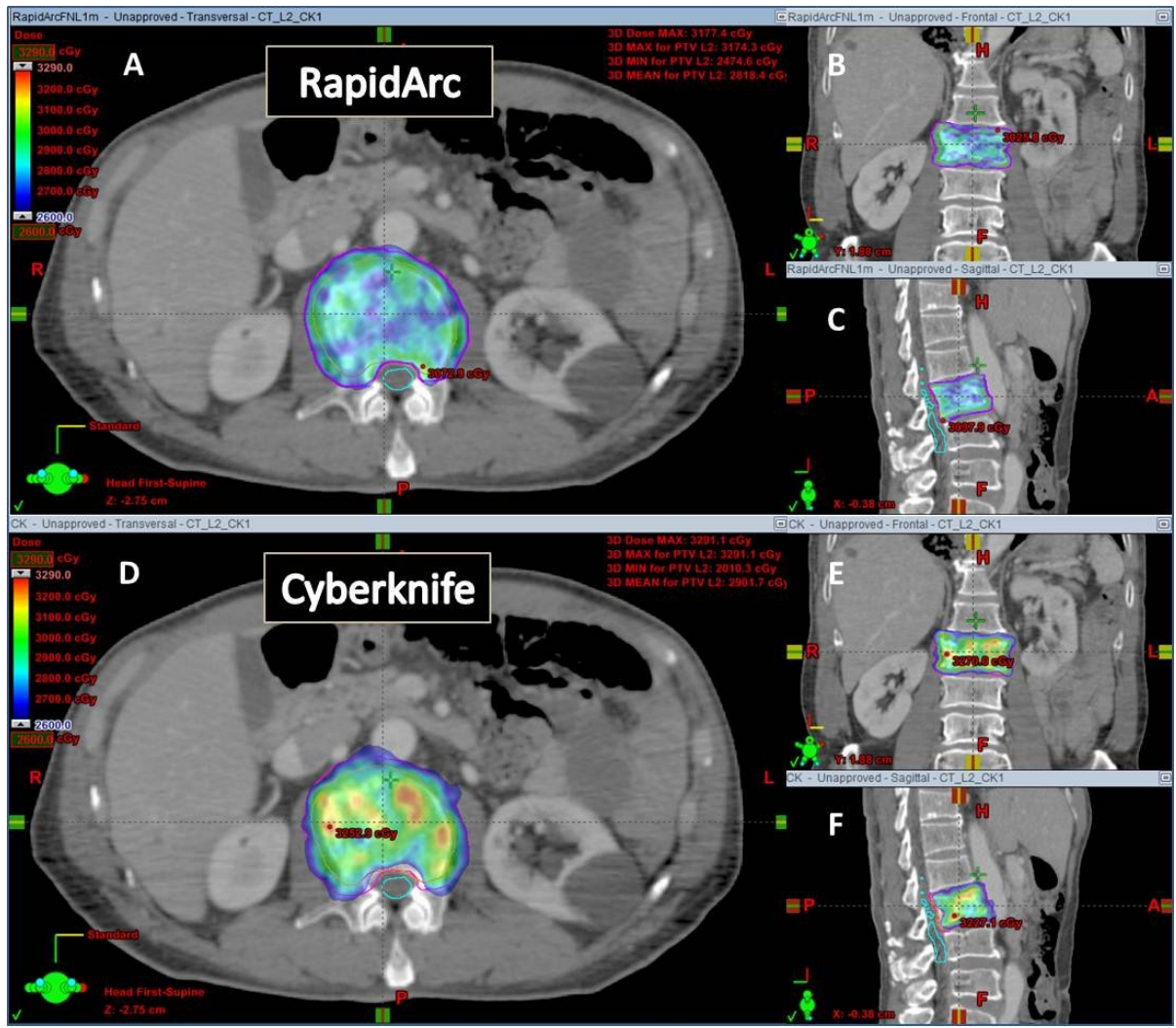

**Figure 3. Dose distribution and PTV coverage of RapidArc vs. CyberKnife in axial (A, D), coronal (B, E), and sagittal (C, F) sections.** The PTV is outlined in pink. RA doses were uniformly distributed inside the PTV and avoided the spinal cord (cyan).

**Table 3. Dosimetric parameter results for the target and organ at risk.**

| Parameter | RapidArc | CyberKnife | p-value |
|---|---|---|---|
| *Conformity* | | | |
| $R_{100\%}$ | 1.11 ± 0.03 | 1.17 ± 0.10 | 0.074 |
| *High-dose spillage* | | | |
|  | 0.86 ±1.30 | 2.26 ± 4.21 | 0.203 |
| *Intermediate-dose spillage* | | | |
| $R_{50\%}$ | 4.61 ± 1.85 | 5.32 ± 0.97 | 0.074 |
| $D_{2cm}$ | 57.42 ± 7.38 | 72.56 ± 16.13 | 0.001* |
| *Spinal cord* | | | |
| $D_{max}$ [cGy] | 1577.57 ± 74.96 | 1909.10 ± 60.95 | 0.000* |

$R_{xx\%}$, ratio of the percentage of the prescribed dose volume/PTV volume; $D_{2cm}$, maximum dose in % of prescription dose at 2 cm beyond the PTV in any direction.

*Statistically significant ($p < 0.05$).

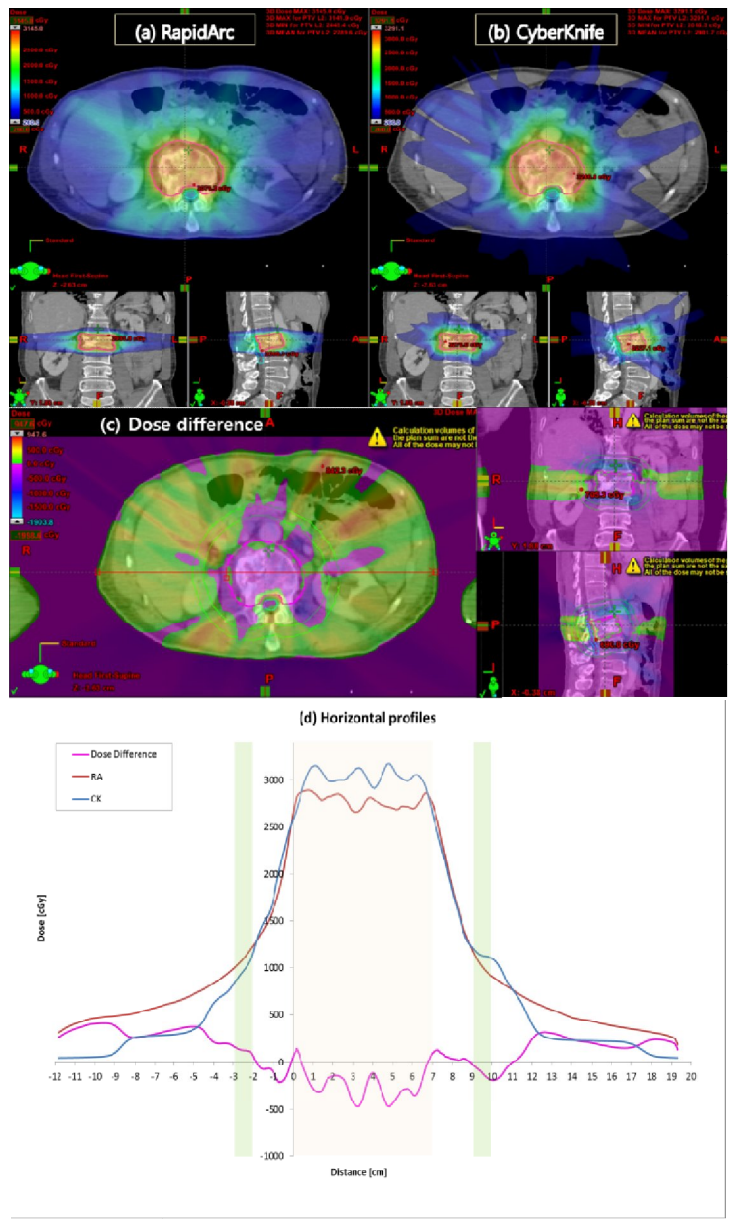

**Figure 4. Representative dose distributions for (a) CK and (b) RA, and (c) differences in the dose distributions of the two plans (RA-CK).** (c) The oblique patterns of the non-coplanar beams are shown for the range of negative values (pink), and dose differences greater than 0 indicate coplanar beams with two complete arcs (green). (d) Horizontal profiles of each dose distribution. The highlighted box from 0–7 cm indicates the PTV area, and two 1 cm areas are highlighted as intermediate-spillage areas.

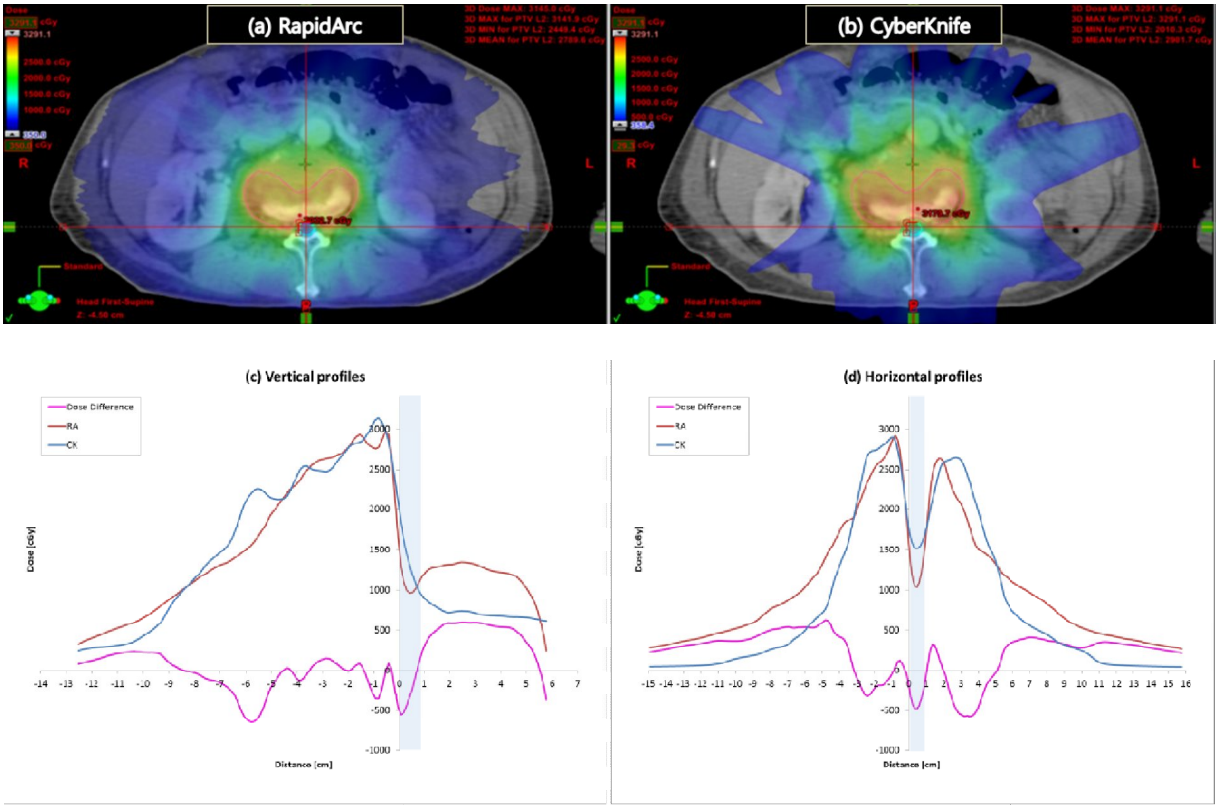

**Figure 5**. **(a) RA and (b) CK dose profiles.** The vertical and horizontal lines are drawn through the center of the spinal cord. The two graphs show the (c) vertical and (d) horizontal dose profiles for RA (red line), CK (blue line), and differences between these two plans (pink line). The spinal cord region is indicated by the highlighted blue box. The RA plan has steeper dose gradients just outside the PTV indicating more efficiently saving the spinal cord.

## IV. Discussion and Conclusion

Conventional RT is a common way to effectively control pain in patients with spinal metastasis. However, in conventional RT, the full dose can result in delivery of radiation to vertebral bodies, including the spinal cord, which are not resistant to high-dose radiation. SBRT is applied to treat spinal tumors and deliver high doses to control tumors, reduce the retreatment rate, and prevent spinal cord compression. SBRT has promising advantages over conventional RT for use in palliation regimens [17]. SBRT has also been performed in combination with various systems that have different characteristics, including robotic linear accelerators. Both CyberKnife and Varian RapidArc are used at our institution. Many facilities use these techniques; however, direct dosimetric comparisons between CK and RA plans have not been reported.

In our current analyses, we directly compared these two systems in terms of plan quality using several dosimetric indices according to RTOG protocols. Although both systems demonstrated good abilities and highly conformal volumetric dose distributions, RA demonstrated lower high- and intermediate-dose spillage than CK. According to the subtracted dose distribution, both plans had sharp dose gradients outside the target; however, as shown in Figure 5, spinal cord regions were more effectively protected with the RA plan, which has a lower maximum dose and steeper dose gradient than the CK plan. All patients treated by CK received a higher maximum dose to the spinal cord in comparison with RA (approximately 11.6% higher). The maximum dose received by the spinal cord may be related to the functions of several parameters, including target volume and the proximity of the spinal cord to the target.

The major features of RA are uniform dose distributions within the PTV and higher conformity. More importantly, it avoids high-dose delivery to the spinal cord, which can be attributed to the use of arc therapy and flattening filter-free beams that provide steep dose fall-off profiles and less dependence on field size. In addition, flattening filter-free beams increase the dose rate by up to 2400 MU/minute, allowing for shorter treatment times, greater cell killing, and more potent biological effects [18]. The high spatial resolution of the multileaf collimator (2.5 mm wide in the central 8 cm at the isocenter) improves conformity in comparison with the more heterogeneous dose distribution of cone-based CK. Similar results were recently reported; specifically, RA was shown to be more attractive than CK in terms of dosimetric distribution and shorter treatment times in patients with prostate metastasis [19].

It should be mentioned that in this study 3-mm PTV margin is applied for the CK treatment, while no PTV margin is recommended by RTOG 0631 assuming no setup error for single fraction radiosurgery treatment [13, 20, 21]. Since 3 or 4 fractionations are used in our institution, 3 mm PTV margin is applied in practice to take into account inter-fractional setup variations. With improved accuracy of image guidance using CBCT, PTV margin for RA would be similar as that of CK.

In conclusion, it appears that RA in combination with two complete coplanar arcs achieves better plan quality, reduces the dose to the spinal cord, and provides adequate target coverage with slightly better CI values than CK, although not every index revealed statistical significance. Therefore, RA for spinal SBRT may demonstrate additional dosimetric advantages that could lead to clinical benefits.

**Acknowledgments**

This work was supported by the Radiation Technology R&D program (2013M2A2A7043506) through the National Research Foundation of Korea, funded by the Ministry of Science, ICT & Future Planning.

# REFERENCES

[1] J.S. Kuo *et al.*, Neurosurgery, **53**, 5 (2003).

[2] N.N. Laack and P.D. Brown, *Cognitive sequelae of brain radiation*, Lippincott Williams & Wilkins, Philadelphia 2011.

[3] E. Chow *et al.*, Journal of Clinical Oncology, **25**, 11 (2007).

[4] A.S. Shiu *et al.*, International Journal of Radiation Oncology Biology Physics, **57**, 3 (2003).

[5] A.K. Garg *et al.*, Cancer, **117**, 15 (2011).

[6] M. Foote *et al.*, Journal of Clinical Neuroscience, **18**, 2 (2011).

[7] R. John *et al.,* Journal of Neurosurgery**, 76**, 3 (1992).

[8] B.L. Guthrie and J.R. Alder, Clinical neurosurgery, **38**,(1992).

[9] M. Murphy and R.S. Cox, Medical Physics, **23**,12 (1996)

[10] J. Hrbacek *et al.,* Medical Physics, **41**, 3 (2014)

[11] K. Otto, Medical Physics, **35**, 1 (2008)

[12] W.F Verbakel *et al.*, Radiotherapy and Oncology, **93**, 1,(2009)

[13] RTOG 0631. Phase II/III Study of Image-Guided Radiosurgery/SBRT for Localized Spine Metastasis.

[14] RTOG 0915. A Randomized Phase II Study Comparing 2 Stereotactic Body Radiation Therapy (SBRT) Schedules for Medically Inoperable Patients with Stage I Peripheral Non-Small Cell Lung Cancer.

[15] L. Feuvret *et al.,* International Journal of Radiation Oncology*Biology*Physics, **64**, 2, (2006)

[16] L. Chakravart and Roy, John Wiley and Sons, (1967)

[17] A. Sahgal, D.A. Larson, and E.L. Chang, International Journal of Radiation Oncology Biology Physics, **71**, 3 (2008).

[18] R.D. Timmerman, Seminars in Radiation Oncology, **18**, 4 (2008).

[19] Y.W. Lin *et al.*, Physica Medica, **30**, 6 (2014).

[20] A.K. Ho *et al.*, Neurosurgery, **60**, 2 (2007)

[21] C. Yu *et al.*, Neurosurgery, **55**, 5 (2004)